\documentclass[prd,nofootinbib,showpacs,preprint]{revtex4}
\usepackage{amsmath}
\usepackage{graphicx}
\usepackage{epsfig}
\graphicspath{{Figs/}}
\usepackage{dcolumn}
\usepackage{bm}
\usepackage{amssymb}
\usepackage[usenames,dvipsnames]{color}
\usepackage{slashed}
\usepackage[dvipdfm,colorlinks,citecolor=blue]{hyperref}
\usepackage{booktabs}
\usepackage{multirow}
\begin{document}
\title{Corrections to Scaling Neutrino Mixing: Non-zero $\theta_{13}, \delta_{CP}$ and Baryon Asymmetry}
\author{Rupam Kalita}
\email{rup@tezu.ernet.in}
\author{Debasish Borah}
\email{dborah@tezu.ernet.in}
\author{Mrinal Kumar Das}
\email{mkdas@tezu.ernet.in}
\affiliation{Department of Physics, Tezpur University, Tezpur-784028, India}

\begin{abstract}
We study a very specific type of neutrino mass and mixing structure based on the idea of Strong Scaling Ansatz (SSA) where the ratios of neutrino mass matrix elements belonging to two different columns are equal. There are three such possibilities, all of which are disfavored by the latest neutrino oscillation data. We focus on the specific scenario which predicts vanishing reactor mixing angle $\theta_{13}$ and inverted hierarchy with vanishing lightest neutrino mass. Motivated by several recent attempts to explain non-zero $\theta_{13}$ by incorporating corrections to a leading order neutrino mass or mixing matrix giving $\theta_{13}=0$, here we study the origin of non-zero $\theta_{13}$ as well as leptonic Dirac CP phase $\delta_{CP}$ by incorporating two different corrections to scaling neutrino mass and mixing: one where type II seesaw acts as a correction to scaling neutrino mass matrix and the other with charged lepton correction to scaling neutrino mixing. Although scaling neutrino mass matrix originating from type I seesaw predicts inverted hierarchy, the total neutrino mass matrix after type II seesaw correction can give rise to either normal or inverted hierarchy. However, charged lepton corrections do not disturb the inverted hierarchy prediction of scaling neutrino mass matrix. We further discriminate between neutrino hierarchies, different choices of lightest neutrino mass and Dirac CP phase by calculating baryon asymmetry and comparing with the observations made by the Planck experiment.

\end{abstract}
\maketitle
\section{Introduction}
\label{intro}
Origin of tiny neutrino masses and mixing is one of the most widely studied problems in modern day particle physics. Since the standard model (SM) of particle physics fails to provide an explanation to neutrino masses and mixing, several well motivated beyond standard model (BSM) frameworks have been proposed to account for the tiny neutrino mass observed by several neutrino oscillation experiments \cite{PDG}. More recently, the neutrino oscillation experiments T2K \cite{T2K}, Double ChooZ \cite{chooz}, Daya-Bay \cite{daya} and RENO \cite{reno} have also confirmed the earlier results and also made the measurement of neutrino parameters more precise. The latest global fit values for $3\sigma$ range of neutrino oscillation parameters \cite{schwetz14} are shown in table \ref{tab:data1}. Another global fit study \cite{valle14} reports the 3$\sigma$ values as shown in table \ref{tab:data2}.
\begin{center}
\begin{table}[htb]
\begin{tabular}{|c|c|c|}
\hline
Parameters & Normal Hierarchy (NH) & Inverted Hierarchy (IH) \\
\hline
$ \frac{\Delta m_{21}^2}{10^{-5} \text{eV}^2}$ & $7.02-8.09$ & $7.02-8.09 $ \\
$ \frac{|\Delta m_{31}^2|}{10^{-3} \text{eV}^2}$ & $2.317-2.607$ & $2.307-2.590 $ \\
$ \sin^2\theta_{12} $ &  $0.270-0.344 $ & $0.270-0.344 $ \\
$ \sin^2\theta_{23} $ & $0.382-0.643$ &  $0.389-0.644 $ \\
$\sin^2\theta_{13} $ & $0.0186-0.0250$ & $0.0188-0.0251 $ \\
$ \delta_{CP} $ & $0-2\pi$ & $0-2\pi$ \\
\hline
\end{tabular}
\caption{Global fit $3\sigma$ values of neutrino oscillation parameters \cite{schwetz14}}
\label{tab:data1}
\end{table}
\end{center}
\begin{center}
\begin{table}[htb]
\begin{tabular}{|c|c|c|}
\hline
Parameters & Normal Hierarchy (NH) & Inverted Hierarchy (IH) \\
\hline
$ \frac{\Delta m_{21}^2}{10^{-5} \text{eV}^2}$ & $7.11-8.18$ & $7.11-8.18 $ \\
$ \frac{|\Delta m_{31}^2|}{10^{-3} \text{eV}^2}$ & $2.30-2.65$ & $2.20-2.54 $ \\
$ \sin^2\theta_{12} $ &  $0.278-0.375 $ & $0.278-0.375 $ \\
$ \sin^2\theta_{23} $ & $0.393-0.643$ &  $0.403-0.640 $ \\
$\sin^2\theta_{13} $ & $0.0190-0.0262$ & $0.0193-0.0265 $ \\
$ \delta_{CP} $ & $0-2\pi$ & $0-2\pi$ \\
\hline
\end{tabular}
\caption{Global fit $3\sigma$ values of neutrino oscillation parameters \cite{valle14}}
\label{tab:data2}
\end{table}
\end{center}
Although the $3\sigma$ range for the Dirac CP phase $\delta_{CP}$ is $0-2\pi$, there are two possible best fit values of it found in the literature: $306^o$ (NH), $254^o$ (IH) \cite{schwetz14} and $254^o$ (NH), $266^o$ (IH) \cite{valle14}. It should be noted that the neutrino oscillation experiments only determine two mass squared differences and hence the lightest neutrino mass is still unknown. Cosmology experiments however puts an upper bound on the sum of absolute neutrino masses $\sum_i \lvert m_i \rvert < 0.23$ eV \cite{Planck13}. Within this bound, he lightest neutrino mass can either be zero or very tiny (compared to the other two) giving rise to a hierarchical pattern. Or, the lightest neutrino mass can be of same order as the other two neutrino masses giving rise to a quasi-degenerate type neutrino mass spectrum.

Apart from the issue of lightest neutrino mass and hence the nature of neutrino mass hierarchy, the CP violation in the leptonic sector is also not understood very well. Non-zero CP violation in the leptonic sector can be very significant from cosmology point of view as it could be the origin of matter-antimatter asymmetry in the Universe. The latest data available from Planck mission constrain the baryon asymmetry \cite{Planck13} as
\begin{equation}
Y_B = (8.58 \pm 0.22) \times 10^{-11}
\label{barasym}
\end{equation} 
Leptogenesis is one of the most promising dynamical mechanism of generating this observed baryon asymmetry in the Universe by generating an asymmetry in the leptonic sector first which later gets converted into baryon asymmetry through $B+L$ violating electroweak sphaleron transitions \cite{sphaleron}. As pointed out first by Fukugita and Yanagida \cite{fukuyana}, the out of equilibrium CP violating decay of heavy Majorana neutrinos provides a natural way to create the required lepton asymmetry. The most notable feature of this mechanism is that it connects two of the most widely studied problems in particle physics: the origin of neutrino mass and the origin of baryon asymmetry. This idea has been explored within several interesting BSM frameworks \cite{leptoreview,joshipura,davidsonPR}. Recently such a comparative study was done to understand the impact of mass hierarchies, Dirac and Majorana CP phases on the predictions for baryon asymmetry in \cite{leptodborah} within the framework of left-right symmetric 
models. 

Motivated by the quest for understanding the origin of neutrino masses and mixing and its relevance in cosmology, we recently studied several models \cite{devTBM} based on the idea of generating non-zero $\theta_{13}$, $\delta_{CP}$ and matter-antimatter asymmetry by perturbing generic $\mu-\tau$ symmetric neutrino mass matrix which can be explained dynamically within generic flavor symmetry models. In these works, type I seesaw \cite{ti} is assumed to give rise to the $\mu-\tau$ symmetric neutrino mass matrix with $\theta_{13}=0$ whereas type II seesaw \cite{tii} acts as a perturbation in order to generate the non-zero reactor mixing angle $\theta_{13}$ and also the Dirac CP phase $\delta_{CP}$ in some cases. In continuation of our earlier works on exploring the underlying structure of the neutrino mass matrix, in this work we consider a very specific neutrino mass matrix structure proposed few years back by the authors of \cite{PLB644}. The structure of the neutrino mass matrix is based on the idea of strong scaling Ansatz where certain ratios of the elements of neutrino mass matrix are equal. Out of three such possibilities (to be discussed in the next section), one of them predicts $\theta_{13}=0$ and an inverted hierarchy with vanishing lightest neutrino mass. Such a scaling neutrino mass matrix can also find its origin in specific flavor symmetry models as discussed in \cite{PLB644}. Several phenomenological studies based on the idea of SSA have appeared in \cite{scalingpheno}. The predictions for neutrino sector similar to the scaling ansatz can also be found in models based on the abelian symmetry $L_e-L_{\mu}-L_{\tau}$ \cite{JPG31}.

Although inverted hierarchy as predicted by SSA can still be viable, vanishing reactor mixing angle is no longer acceptable after the discovery of non-zero $\theta_{13}$. Generation of non-zero $\theta_{13}$ in models based on the idea of SSA have appeared recently in \cite{theta13scaling}. In this work we study two different possibilities of generating non-zero $\theta_{13}$ as well as Dirac CP phase $\delta_{CP}$ by incorporating corrections to either the neutrino mass matrix or the leptonic mixing matrix, also known as the Pontecorvo-Maki-Nakagawa-Sakata (PMNS) matrix. In both the cases we assume the origin of scaling neutrino mass matrix in type I seesaw. The required deviation from scaling can either come from a different seesaw mechanism (say, type II seesaw) or from charged lepton (CL) correction. The crucial difference between the two different scenario is that in CL correction, the inverted hierarchy prediction of SSA remains intact whereas with the combination of two different seesaw mechanism both normal and inverted hierarchies can emerge out of the total neutrino mass matrix. We first numerically fit the scaling neutrino mass matrix (from type I seesaw) with neutrino data on two mass squared differences and two angles $\theta_{12}, \theta_{23}$ (as $\theta_{13}=0$). Then we derive the necessary perturbation to scaling neutrino mixing by comparing with the full neutrino oscillation data including non-zero $\theta_{13}$. We further constrain the perturbation by demanding successful production of baryon asymmetry through the mechanism of leptogenesis.

This paper is organized as follows. In section \ref{sec:scaling}, we briefly discuss the idea of scaling neutrino mass and mixing. In section \ref{sec:devscaling}, we study the possible deviation from scaling with type II seesaw as well as charged lepton corrections. In section \ref{sec:lepto}, we briefly outline the idea of leptogenesis and in section \ref{sec:numeric} we discuss our numerical analysis. We finally conclude in section \ref{sec:conclude}.

\section{Strong Scaling Ansatz}
\label{sec:scaling}
According to SSA, ratios of certain elements of the neutrino mass matrix are equal. The stability of such a structure is also guaranteed by the fact that it is not affected by the renormalization group evolution (RGE) equations. Therefore, the scaling which is present in the neutrino mass matrix at seesaw scale is also remains valid at the weak scale as we run the neutrino parameters from seesaw to weak scale under RGE. We denote the neutrino mass matrix and the leptonic mixing matrix $U_{\text{PMNS}}$ as 
\[
M_{\nu}=
  \begin{pmatrix}
   m_{ee} &  m_{e{\mu}} & m_{e{\tau}}  \\      m_{{\mu}e}  & m_{{\mu}{\mu}} &  m_{{\mu}{\tau}} \\m_{{\tau}e} & m_{{\tau}{\mu}} &     m_{{\tau}{\tau}} \end{pmatrix}
\]

\[
U_{\text{PMNS}} = 
\begin{pmatrix}
U_{e1} & U_{e2} & U_{e3} \\ 
U_{\mu 1} & U_{\mu 2} & U_{\mu 3} \\
U_{\tau 1} & U_{\tau 2} & U_{\tau 3} 
\end{pmatrix}
\]
As noted by the authors of \cite{PLB644}, there are three different types of SSA which can be written as 
\begin{equation}
\label{eq:someequation}
\frac{m_{e{\mu}}}{ m_{e{\tau}}}=\frac{ m_{{\mu}{\mu}}}{ m_{{\mu}{\tau}}} =\frac{ m_{{\tau}{\mu}}}{ m_{{\tau}{\tau}}}=S
\end{equation} 
\begin{equation}
 \label{eq:someequation1}   
\frac{m_{ee}}{ m_{e{\tau}}}=\frac{ m_{{\mu}e}}{ m_{{\mu}{\tau}}} =\frac{ m_{{\tau}e}}{ m_{{\tau}{\tau}}}=S^\prime
\end{equation} 
\begin{equation}
   \label{eq:someequation2}
\frac{m_{ee}}{ m_{e{\mu}}}=\frac{ m_{{\mu}e}}{ m_{{\mu}{\mu}}} =\frac{ m_{{\tau}e}}{ m_{{\tau}{\mu}}}=S^{\prime\prime}
\end{equation} 
Using \eqref{eq:someequation}, we can write the neutrino mass matrix as
\begin{equation}
\label{eq:someequation3}
M_{\nu}=\left(\begin{array}{ccc} A & B & \frac{B}{S}\\
  B & D &\frac{D}{S}\\
\frac{B}{S} &\frac{D}{S} &\frac{D}{S^{2}}\end{array}\right),
\end{equation}
Similarly, for the other two cases \eqref{eq:someequation1}, \eqref{eq:someequation2} one can write down the neutrino mass matrix as 
\begin{equation}
\label{eq:someequation4}
M_{\nu}=\left(\begin{array}{ccc}  A & B & \frac{A}{S^\prime}\\
   B & D &\frac{B}{S^\prime}\\
\frac{A}{S^\prime} &\frac{B}{S^\prime} &\frac{A}{S^{\prime2}} \end{array}\right),
\end{equation}
and
\begin{equation}
\label{eq:someequation5}
M_{\nu}=\left(\begin{array}{ccc}  A & \frac{A}{S^{\prime\prime}} & B\\
   \frac{A}{S^{\prime\prime}} &  \frac{A}{S^{\prime\prime 2}} & \frac{B}{S^{\prime\prime}}\\
 B &\frac{B}{S^{\prime\prime}} &D  \end{array}\right),
\end{equation}
One interesting property of the first scaling mass matrix \eqref{eq:someequation3} is that it has one of its eigenvalue $m_3$ zero (rank 2 matrix) and diagonalization of this matrix gives $ U_{e3}=0$. Thus, it gives rise to inverted hierarchy of neutrino mass with $\theta_{13} = 0$. Although such a scenario is now ruled out after the discovery of non-zero $\theta_{13}$, there still exists the possibility of generating non-zero $\theta_{13}$ by adding perturbations to the scaling neutrino mass and mixing, given the fact that $\theta_{13}$ is still small compared to the other two mixing angles. However, diagonalization of the second scaling matrix \eqref{eq:someequation4} gives  $ U_{\mu3}=0$ or $ U_{\mu1}=0$ depending on the hierarchy of neutrino masses. Similarly, diagonalization of the third scaling matrix \eqref{eq:someequation5} gives  $ U_{\tau3}=0$ or $ U_{\tau1}=0$. The predictions of both the scaling mass matrices obtained using \eqref{eq:someequation1} and \eqref{eq:someequation2} are not phenomenologically viable. Even if we assume the validity of these two scaling mass matrices at tree level, they will require large corrections in order to generate the correct mixing matrix. Leaving these to future studies, here we focus on the possibility of generating non-zero $U_{e3}$ and hence non-zero $\theta_{13}$ by incorporating different corrections to leading order scaling neutrino mass matrix given by \eqref{eq:someequation3}.
\begin{table}[h]
\caption{$M_{RR}$ (in GeV) for Type II seesaw correction  with $\delta_{CP}=\frac{\pi}{2}$}
\begin{center}

\resizebox{\textwidth}{!}{%
\begin{tabular}{ |c| c| c|  }
\hline
\textbf{MODEL} & \textbf{IH $m_3=0.065$ eV}\\
\hline

\addlinespace[1.5ex]
\hline
 \text{$1$ Flavor} &
$ \begin{pmatrix}6.65934\times10^{12}&3.13846\times10^{12} - 1.38275\times10^{12} i&3.66288\times10^{12} - 1.66362\times10^{12} i\\3.13846\times10^{12} - 1.38275\times10^{12} i &7.76821\times10^{12} - 3.14624\times10^{9} i&-5.74326\times10^{11} - 5.85132\times10^{8} i\\3.66288\times10^{12} - 1.66362\times10^{12} i &-5.74326\times10^{11} - 5.85132\times10^{8} i &7.5331\times10^{12} + 3.14624\times10^{9} i \end{pmatrix}$\\
\hline
\addlinespace[2.5ex]
\hline
  \text{$2$ Flavor} &
$ \begin{pmatrix}1.9978\times10^{12}& 9.41538\times10^{11} - 4.14824\times10^{11} i&1.09886\times10^{12} - 4.99085\times10^{11} i\\  9.41538\times10^{11} - 4.14824\times10^{11} i& 2.33046\times10^{12} - 9.43872\times10^{8} i &-1.72298\times10^{11} - 1.7554\times10^{8} i\\1.09886\times10^{12} - 4.99085\times10^{11} i &-1.72298\times10^{11} - 1.7554\times10^{8} i&2.25993\times10^{12} + 9.43872\times10^{8} i\end{pmatrix}$\\
\hline
\addlinespace[2.5ex]
\hline
 \text{$3$ Flavor} &
$ \begin{pmatrix}6.65934\times10^{8}&3.13846\times10^{8} - 1.38275\times10^{8} i &3.66288\times10^{8} - 1.66362\times10^{8} i\\3.13846\times10^{8} - 1.38275\times10^{8} i&7.76821\times10^{8} - 3.14624\times10^{5} i&-5.74326\times10^{7} - 5.85132\times10^{4} i\\3.66288\times10^{8} - 1.66362\times10^{8} i &-5.74326\times10^{7} - 5.85132\times10^{4} i&7.5331\times10^{8} + 3.14624\times10^{5} i\end{pmatrix}$\\
\hline
\end{tabular}
}
\end{center}
\label{table:MRRT21}
\end{table}
\begin{table}[h]
\caption{$M_{RR}$ (in GeV) for Type II seesaw correction  with $\delta_{CP}=\frac{\pi}{2}$}
\begin{center}

\resizebox{\textwidth}{!}{%
\begin{tabular}{ |c| c| c|  }
\hline
\textbf{MODEL} & \textbf{IH $m_3=10^{-6}$ eV}\\
\hline

\addlinespace[1.5ex]
\hline
 \text{$1$ Flavor} &
$ \begin{pmatrix}3.66324\times10^{12}&3.14907\times10^{12} - 4.63416\times10^{11} i &3.65406\times10^{12} - 5.57547\times10^{11} i\\3.14907\times10^{12} - 4.63416\times10^{11} i &3.3088\times10^{12} - 5.22573\times10^{9} i&-2.008\times10^{12} - 9.71871\times10^{8} i\\3.65406\times10^{12} - 5.57547\times10^{11} i &-2.008\times10^{12} - 9.71871\times10^{8} i &2.54042\times10^{12} + 5.22573\times10^{9} i \end{pmatrix}$\\
\hline
\addlinespace[2.5ex]
\hline
  \text{$2$ Flavor} &
$ \begin{pmatrix}1.09897\times10^{12}& 9.44722\times10^{11} - 1.39025\times10^{11} i&1.09622\times10^{12} - 1.67264\times10^{11} i\\  9.44722\times10^{11} - 1.39025\times10^{11} i& 9.92639\times10^{11} - 1.56772\times10^{9} i &-6.02401\times10^{11} - 2.91561\times10^{8} i\\1.09622\times10^{12} - 1.67264\times10^{11} i &-6.02401\times10^{11} - 2.91561\times10^{8}  i&7.62125\times10^{11} + 1.56772\times10^{9} i\end{pmatrix}$\\
\hline
\addlinespace[2.5ex]
\hline
 \text{$3$ Flavor} &
$ \begin{pmatrix}3.66324\times10^{8}&3.14907\times10^{8} - 4.63416\times10^{7} i &3.65406\times10^{8} - 5.57547\times10^{7} i\\3.14907\times10^{8} - 4.63416\times10^{7} i&3.3088\times10^{8} - 5.22573\times10^{5} i&-2.008\times10^{8} - 9.71871\times10^{4} i\\3.65406\times10^{8} - 5.57547\times10^{7} i &-2.008\times10^{8} - 9.71871\times10^{4} i &2.54042\times10^{8} + 5.22573\times10^{5} i\end{pmatrix}$\\
\hline
\end{tabular}
}
\end{center}
\label{table:MRRT22}
\end{table}
\begin{table}[h]
\caption{$M_{RR}$ (in GeV) for Type II seesaw correction  with $\delta_{CP}=\frac{\pi}{2}$}
\begin{center}

\resizebox{\textwidth}{!}{%
\begin{tabular}{ |c| c| c|  }
\hline
\textbf{MODEL} & \textbf{NH $m_1=0.07$ eV}\\
\hline

\addlinespace[1.5ex]
\hline
 \text{$1$ Flavor} &
$ \begin{pmatrix}5.53224\times10^{12}&3.14101\times10^{12} - 1.47308\times10^{12} i&3.66076\times10^{12} - 1.7723\times10^{12} i\\3.14101\times10^{12} - 1.47308\times10^{12} i&7.93362\times10^{12} - 3.64514\times10^9 i&9.77698\times10^{11} - 6.77917\times10^8 i\\3.66076\times10^{12} - 1.7723\times10^{12} i&9.77698\times10^{11} - 6.77917\times10^8 i&8.27579\times10^{12} + 3.64514\times10^9 i\end{pmatrix}$\\
\hline
\addlinespace[2.5ex]
\hline
  \text{$2$ Flavor} &
$ \begin{pmatrix}1.65967\times10^{12}&9.42302\times10^{11} - 4.41923\times10^{11} i&1.09823\times10^{12} - 5.31689\times10^{11} i\\9.42302\times10^{11} - 4.41923\times10^{11} i&2.38008\times10^{12} - 1.09354\times10^9 i &2.93309\times10^{11} - 2.03375\times10^8 i\\1.09823\times10^{12} - 5.31689\times10^{11} i&2.93309\times10^{11} - 2.03375\times10^8 i&2.48274\times10^{12} + 1.09354\times10^9 i\end{pmatrix}$\\
\hline
\addlinespace[2.5ex]
\hline
 \text{$3$ Flavor} &
$ \begin{pmatrix}5.53224\times10^8&3.14101\times10^8 - 1.47308\times10^8 i&3.66076\times10^8 - 1.7723\times10^8 i\\3.14101\times10^8 - 1.47308\times10^8 i&7.93362\times10^8 - 364514. i&9.77698\times10^7 - 67791.7 i\\3.66076\times10^8 - 1.7723\times10^8 i&9.77698\times10^7 - 67791.7 i& 8.27579\times10^8 + 364514.1 i\end{pmatrix}$\\
\hline
\end{tabular}
}
\end{center}
\label{table:MRRT23}
\end{table}
\begin{table}[h]
\caption{$M_{RR}$ (in GeV) for Type II seesaw correction  with $\delta_{CP}=\frac{\pi}{2}$}
\begin{center}

\resizebox{\textwidth}{!}{%
\begin{tabular}{ |c| c| c|  }
\hline
\textbf{MODEL} & \textbf{NH $m_1=10^{-6}$ eV}\\
\hline

\addlinespace[1.5ex]
\hline
 \text{$1$ Flavor} &
$ \begin{pmatrix}-9.27517\times10^{12}&3.42428\times10^{13} -4.92222\times10^{12} i&3.42531\times10^{13} - 5.92205\times10^{12} i\\3.42428\times10^{13} -4.92222\times10^{12} i&2.73041\times10^{13} - 5.91445\times10^{11} i&2.53726\times10^{13} - 1.09996\times10^{11} i\\3.42531\times10^{13} - 5.92205\times10^{12} i&2.53726\times10^{13} - 1.09996\times10^{11}  i&3.65267\times10^{13} + 5.91445\times10^{11} i\end{pmatrix}$\\
\hline
\addlinespace[2.5ex]
\hline
  \text{$2$ Flavor} &
$ \begin{pmatrix}-2.78255\times10^{11}&1.02729\times10^{12} - 1.47667\times10^{11} i&1.02759\times10^{12} - 1.77661\times10^{11} i\\1.02729\times10^{12} - 1.47667\times10^{11} i&8.19122\times10^{11} - 1.77434\times10^{10} i &7.61178\times10^{11} - 3.29987\times10^9 i\\1.02759\times10^{12} - 1.77661\times10^{11} i&7.61178\times10^{11} - 3.29987\times10^9  i&1.0958\times10^{12} + 1.77434\times10^{10} i\end{pmatrix}$\\
\hline
\addlinespace[2.5ex]
\hline
 \text{$3$ Flavor} &
$ \begin{pmatrix}-9.27517\times10^7&3.42428\times10^8 - 4.92222\times10^7 i&3.42531\times10^8 - 5.92205\times10^7 i\\3.42428\times10^8 - 4.92222\times10^7 i&2.73041\times10^8 - 5.91445\times10^6  i&2.53726\times10^8 - 1.09996\times10^6 i\\3.42531\times10^8 - 5.92205\times10^7 i&2.53726\times10^8 - 1.09996\times10^6 i& 3.65267\times10^8 + 5.91445\times10^6 i\end{pmatrix}$\\
\hline
\end{tabular}
}
\end{center}
\label{table:MRRT24}
\end{table}
\begin{table}[h]
\caption{$m_{LR}$ (in GeV) for Type II seesaw correction with $\delta_{CP}=\frac{\pi}{2}$}
\begin{center}

\resizebox{\textwidth}{!}{%
\begin{tabular}{ |c| c| c|  }
\hline
\textbf{$a=b$} & \textbf{IH $m_3=0.065$ eV}\\
\hline

\addlinespace[1.5ex]
\hline
 \text{$1$ Flavor} &
$ \begin{pmatrix}5.99679 - 0.179954 i &5.99679 - 0.179954 i&7.06046 - 0.211873 i\\5.99679 - 0.179954 i&20.0543 - 7.83146 i&-23.6114 - 9.22054 i\\7.06046 - 0.211873 i&-23.6114 - 9.22054 i&-27.7994 - 10.856 i \end{pmatrix}$\\
\hline
\addlinespace[2.5ex]
\hline
  \text{$2$ Flavor} &
$ \begin{pmatrix} 3.28458 - 0.098565 i& 3.28458 - 0.098565 i&3.86717 - 0.116048 i\\3.28458 - 0.098565 i&-10.9842 - 4.28947 i &-12.9325 - 5.0503 i\\3.86717 - 0.116048 i&-12.9325 - 5.0503 i&-15.2264 - 5.94608 i\end{pmatrix}$\\
\hline
\addlinespace[2.5ex]
\hline
 \text{$3$ Flavor} &
$ \begin{pmatrix}0.0599679 - 0.00179954 i&0.0599679 - 0.00179954 i&0.0706046 - 0.00211873 i\\0.0599679 - 0.00179954 i&-0.200543 - 0.0783146 i &-0.236114 - 0.0922054 i\\0.0706046 - 0.00211873 i&-0.236114 - 0.0922054 i&-0.277994 - 0.10856 i\end{pmatrix}$\\
\hline
\end{tabular}
}
\end{center}
\label{table:mLRT21}
\end{table}
\begin{table}[h]
\caption{$m_{LR}$ (in GeV) for Type II seesaw correction with $\delta_{CP}=\frac{\pi}{2}$}
\begin{center}

\resizebox{\textwidth}{!}{%
\begin{tabular}{ |c| c| c|  }
\hline
\textbf{$a=b$} & \textbf{IH $m_3=10^{-6}$ eV}\\
\hline

\addlinespace[1.5ex]
\hline
 \text{$1$ Flavor} &
$ \begin{pmatrix}6.20838 + 0.177014 i &6.20838 + 0.177014 i&7.30957 + 0.208411 i\\6.20838 + 0.177014 i&44.2529 - 72.3072 i&52.1021 - 85.1326 i\\7.30957 + 0.208411 i&52.1021 - 85.1326 i&61.3436 - 100.233 i \end{pmatrix}$\\
\hline
\addlinespace[2.5ex]
\hline
  \text{$2$ Flavor} &
$ \begin{pmatrix}3.40047 + 0.0969544 i& 3.40047 + 0.0969544 i  &4.00362 + 0.114151 i\\3.40047 + 0.0969544 i  &24.2383 - 39.6043 i &28.5375 - 46.629 i\\4.00362 + 0.114151 i&28.5375 - 46.629 i&33.5993 - 54.8997 i\end{pmatrix}$\\
\hline
\addlinespace[2.5ex]
\hline
 \text{$3$ Flavor} &
$ \begin{pmatrix}0.0620838 + 0.00177014 i&0.0620838 + 0.00177014 i &0.0730957 + 0.00208411 i\\0.0620838 + 0.00177014 i&0.442529 - 0.723072 i&0.521021 - 0.851326 i\\0.0730957 + 0.00208411 i&0.521021 - 0.851326 i& 0.613436 - 1.00233 i\end{pmatrix}$\\
\hline
\end{tabular}
}
\end{center}
\label{table:mLRT22}
\end{table}
\begin{table}[h]
\caption{$m_{LR}$ (in GeV) for Type II seesaw correction with $\delta_{CP}=\frac{\pi}{2}$}
\begin{center}

\resizebox{\textwidth}{!}{%
\begin{tabular}{ |c| c| c|  }
\hline
\textbf{$b=d$} & \textbf{IH $m_3=0.065$ eV}\\
\hline

\addlinespace[1.5ex]
\hline
 \text{$1$ Flavor} &
$ \begin{pmatrix}-64.8942 - 43.2459 i&-0.112269 - 3.74126 i&-0.132183 - 4.40486 i\\-0.112269 - 3.74126 i &-0.112269 - 3.74126 i&-0.132183 - 4.40486 i\\-0.132183 - 4.40486  i&-0.132183 - 4.40486 i &-0.155628 - 5.18616 i \end{pmatrix}$\\
\hline
\addlinespace[2.5ex]
\hline
  \text{$2$ Flavor} &
$ \begin{pmatrix} -35.544 - 23.6868 i& -0.0614925 - 2.04917 i &-0.0723996 - 2.41264 i\\-0.0614925 - 2.04917 i &-0.0614925 - 2.04917 i &-0.0723996 - 2.41264 i\\-0.0723996 - 2.41264  i&-0.0723996 - 2.41264 i&-0.0852412 - 2.84058 i\end{pmatrix}$\\
\hline
\addlinespace[2.5ex]
\hline
 \text{$3$ Flavor} &
$ \begin{pmatrix}-0.648942 - 0.432459 i&-0.00112269 - 0.0374126 i &-0.00132183 - 0.0440486 i\\-0.00112269 - 0.0374126 i&-0.00112269 - 0.0374126  i&-0.00132183 - 0.0440486 i\\-0.00132183 - 0.0440486 i &-0.00132183 - 0.0440486 i& -0.00155628 - 0.0518616 i\end{pmatrix}$\\
\hline
\end{tabular}
}
\end{center}
\label{table:mLRT23}
\end{table}
\begin{table}[h]
\caption{$m_{LR}$ (in GeV) for Type II seesaw correction  with $\delta_{CP}=\frac{\pi}{2}$}
\begin{center}

\resizebox{\textwidth}{!}{%
\begin{tabular}{ |c| c| c|  }
\hline
\textbf{$b=d$} & \textbf{IH $m_3=10^{-6}$ eV}\\
\hline

\addlinespace[1.5ex]
\hline
 \text{$1$ Flavor} &
$ \begin{pmatrix}-4.28715 - 25.1506 i&0.110435 - 3.87326 i &0.130023 - 4.56027 i\\0.110435 - 3.87326 i &0.110435 - 3.87326 i&0.130023 - 4.56027 i\\0.130023 - 4.56027 i &0.130023 - 4.56027  i &0.153085 - 5.36914 i \end{pmatrix}$\\
\hline
\addlinespace[2.5ex]
\hline
  \text{$2$ Flavor} &
$ \begin{pmatrix}-2.34817 - 13.7755 i& 0.0604877 - 2.12147 i &0.0712165 - 2.49777 i\\ 0.0604877 - 2.12147 i& 0.0604877 - 2.12147 i &0.0712165 - 2.49777 i\\0.0712165 - 2.49777 i &0.0712165 - 2.49777 i&0.0838483 - 2.9408 i\end{pmatrix}$\\
\hline
\addlinespace[2.5ex]
\hline
 \text{$3$ Flavor} &
$ \begin{pmatrix}0.0428715 + 0.251506 i&-0.00110435 + 0.0387326 i &-0.00130023 + 0.0456027 i\\-0.00110435 + 0.0387326 i&-0.00130023 + 0.0456027 i&-0.00130023 + 0.0456027 i\\-0.00130023 + 0.0456027 i &-0.00130023 + 0.0456027 i&-0.00153085 + 0.0536914 i\end{pmatrix}$\\
\hline
\end{tabular}
}
\end{center}
\label{table:mLRT24}
\end{table}
\begin{table}[h]
\caption{$m_{LR}$ (in GeV) for Type II seesaw correction  with $\delta_{LR}=\frac{\pi}{2}$}
\begin{center}

\resizebox{\textwidth}{!}{%
\begin{tabular}{ |c| c| c|  }
\hline
\textbf{$a=b$} & \textbf{NH $m_1=0.07$ eV}\\
\hline

\addlinespace[1.5ex]
\hline
 \text{$1$ Flavor} &
$ \begin{pmatrix}6.36821 - 0.52521 i &6.36821 - 0.52521 i&7.49775 - 0.618368 i\\6.36821 - 0.52521 i&-31.387 - 5.45116 i&-36.9542 - 6.41805 i\\7.49775 - 0.618368 i&-36.9542 - 6.41805 i&-43.5088 - 7.55643 i \end{pmatrix}$\\
\hline
\addlinespace[2.5ex]
\hline
  \text{$2$ Flavor} &
$ \begin{pmatrix} 3.48801 - 0.28767 i& 3.48801 - 0.28767 i&4.10669 - 0.338694 i\\3.48801 - 0.28767 i&-17.1914 - 2.98572 i &-20.2406 - 3.51531 i\\4.10669 - 0.338694 i&-20.2406 - 3.51531 i&-23.8308 - 4.13883 i\end{pmatrix}$\\
\hline
\addlinespace[2.5ex]
\hline
 \text{$3$ Flavor} &
$ \begin{pmatrix}0.0636821 - 0.0052521 i&0.0636821 - 0.0052521 i&0.0749775 - 0.00618368 i\\0.0636821 - 0.0052521 i&-0.31387 - 0.0545116 i &-0.369542 - 0.0641805 i\\0.0749775 - 0.00618368 i&-0.369542 - 0.0641805 i& -0.435088 - 0.0755643 i\end{pmatrix}$\\
\hline
\end{tabular}
}
\end{center}
\label{table:mLRT25}
\end{table}
\begin{table}[h]
\caption{$m_{LR}$ (in GeV) for Type II seesaw correction  with $\delta_{CP}=\frac{\pi}{2}$}
\begin{center}

\resizebox{\textwidth}{!}{%
\begin{tabular}{ |c| c| c|  }
\hline
\textbf{$a=b$} & \textbf{NH $m_1=10^{-6}$ eV}\\
\hline

\addlinespace[1.5ex]
\hline
 \text{$1$ Flavor} &
$ \begin{pmatrix}16.3451 - 0.40342 i&16.3451 - 0.40342 i&19.2442 - 0.474975 i\\16.3451 - 0.40342  i&-59.0353 + 7.20559 i&-69.5065 + 8.48366 i\\19.2442 - 0.474975 i&-69.5065 + 8.48366 i&-81.8351 + 9.98843  i \end{pmatrix}$\\
\hline
\addlinespace[2.5ex]
\hline
  \text{$2$ Flavor} &
$ \begin{pmatrix} 2.83105 - 0.0698744 i& 2.83105 - 0.0698744 i&3.33319 - 0.0822681 i\\2.83105 - 0.0698744  i&-10.2252 + 1.24804 i &-12.0389 + 1.46941 i\\3.33319 - 0.0822681 i&-12.0389 + 1.46941  i&-14.1742 + 1.73005 i\end{pmatrix}$\\
\hline
\addlinespace[2.5ex]
\hline
 \text{$3$ Flavor} &
$ \begin{pmatrix}0.0516876 - 0.00127573 i&0.0516876 - 0.00127573 i&0.0608555 - 0.001502 i\\0.0516876 - 0.00127573 i&-0.186686 + 0.0227861 i&-0.219799 + 0.0268277 i\\0.0608555 - 0.001502 i&-0.219799 + 0.0268277 i& -0.258785 + 0.0315862 i\end{pmatrix}$\\
\hline
\end{tabular}
}
\end{center}
\label{table:mLRT26}
\end{table}
\begin{table}[h]
\caption{$m_{LR}$ (in GeV) for Type II seesaw correction  with $\delta_{CP}=\frac{\pi}{2}$}
\begin{center}

\resizebox{\textwidth}{!}{%
\begin{tabular}{ |c| c| c|  }
\hline
\textbf{$b=d$} & \textbf{NH $m_1=0.07$ eV}\\
\hline

\addlinespace[1.5ex]
\hline
 \text{$1$ Flavor} &
$ \begin{pmatrix}-30.1242 - 65.0884 i&0.327667 - 3.97298 i&-0.385786 - 4.67768 i\\0.327667 - 3.97298 i&0.327667 - 3.97298 i&-0.385786 - 4.67768 i\\-0.385786 - 4.67768 i&-0.385786 - 4.67768 i&-0.454214 - 5.50737 i \end{pmatrix}$\\
\hline
\addlinespace[2.5ex]
\hline
  \text{$2$ Flavor} &
$ \begin{pmatrix}16.4997 + 35.6504 i&0.17947 + 2.17609 i&0.211304 + 2.56207 i\\0.17947 + 2.17609 i&0.17947 + 2.17609 i &0.211304 + 2.56207 i\\0.211304 + 2.56207 i&0.211304 + 2.56207 i&0.248783 + 3.01651 i\end{pmatrix}$\\
\hline
\addlinespace[2.5ex]
\hline
 \text{$3$ Flavor} &
$ \begin{pmatrix}-0.301242 - 0.650884 i&-0.00327667 - 0.0397298 i&-0.00385786 - 0.0467768 i\\-0.00327667 - 0.0397298 i&-0.00327667 - 0.0397298 i&-0.00385786 - 0.0467768 i\\-0.00385786 - 0.0467768  i&-0.00385786 - 0.0467768 i& -0.00454214 - 0.0550737 i\end{pmatrix}$\\
\hline
\end{tabular}
}
\end{center}
\label{table:mLRT27}
\end{table}
\begin{table}[h]
\caption{$m_{LR}$ (in GeV) for Type II seesaw correction  with $\delta_{CP}=\frac{\pi}{2}$}
\begin{center}

\resizebox{\textwidth}{!}{%
\begin{tabular}{ |c| c| c|  }
\hline
\textbf{$b=d$} & \textbf{NH $m_1=10^{-6}$ eV}\\
\hline

\addlinespace[1.5ex]
\hline
 \text{$1$ Flavor} &
$ \begin{pmatrix}141.819 - 379.178 i&-0.251684 - 10.1973 i&-0.296326 - 12.006 i\\-0.251684 - 10.1973  i&-0.251684 - 10.1973 i&-0.296326 - 12.006 i\\-0.296326 - 12.006 i&-0.296326 - 12.006 i&-0.348886 - 14.1356 i \end{pmatrix}$\\
\hline
\addlinespace[2.5ex]
\hline
  \text{$2$ Flavor} &
$ \begin{pmatrix}24.5637 - 65.6755 i&-0.043593 - 1.76623 i&-0.0513252 - 2.0795 i\\-0.043593 - 1.76623  i&-0.043593 - 1.76623 i &-0.0513252 - 2.0795 i\\-0.0513252 - 2.0795 i&-0.0513252 - 2.0795 i&-0.0604289 - 2.44835 i\end{pmatrix}$\\
\hline
\addlinespace[2.5ex]
\hline
 \text{$3$ Flavor} &
$ \begin{pmatrix}0.44847 - 1.19906 i&-0.000795896 - 0.0322467 i&-0.000937066 - 0.0379664 i\\-0.000795896 - 0.0322467 i&-0.000795896 - 0.0322467 i&-0.000937066 - 0.0379664 i\\-0.000937066 - 0.0379664 i&-0.000937066 - 0.0379664 i& -0.00110328 - 0.0447006 i\end{pmatrix}$\\
\hline
\end{tabular}
}
\end{center}
\label{table:mLRT28}
\end{table}

\section{Deviations from Scaling}
\label{sec:devscaling}
As discussed in the previous section, the neutrino mass matrices based on the idea of SSA do not give rise to the correct neutrino mixing pattern. Therefore, the scaling neutrino mass or mixing matrix has to be corrected in order to have agreement with neutrino oscillation data. Here we consider two different sources of such corrections to SSA which not only can give rise to correct neutrino mixing but also have different predictions for neutrino mass hierarchy, leptonic Dirac CP phase as well as baryon asymmetry.
\subsection{Deviation from Scaling with Type II Seesaw}
Type II seesaw mechanism is the extension of the standard model with a scalar field $\Delta_L$ which transforms like a triplet under $SU(2)_L$ and has $U(1)_Y$ charge twice that of lepton doublets. Such a choice of gauge structure allows an additional Yukawa term in the Lagrangian given by $ f_{ij}\ \left(\ell_{iL}^T \ C \ i \sigma_2 \Delta_L \ell_{jL}\right)$. The triplet can be represented as 
\begin{equation}
\Delta_L =
\left(\begin{array}{cc}
\ \delta^+_L/\surd 2 & \delta^{++}_L \\
\ \delta^0_L & -\delta^+_L/\surd 2
\end{array}\right) \nonumber
\end{equation} 
The scalar Lagrangian of the standard model also gets modified after the inclusion of this triplet. Apart from the bilinear and quartic coupling terms of the triplet, there is one trilinear term as well involving the triplet and the standard model Higgs doublet. From the minimization of the scalar potential, the neutral component of the triplet is found to acquire a vacuum expectation value (vev) given by 
\begin{equation}
 \langle \delta^0_L \rangle = v_L = \frac{\mu_{\Delta H}\langle \phi^0 \rangle^2}{M^2_{\Delta}}
\label{vev} 
\end{equation}
where $\phi^0=v$ is the neutral component of the electroweak Higgs doublet with vev approximately $10^2$ GeV. The trilinear coupling term $\mu_{\Delta H}$ and the mass term of the triplet $M_{\Delta}$ can be taken to be of same order. Thus, $M_{\Delta}$ has to be as high as $10^{14}$ GeV to give rise to tiny neutrino masses without any fine-tuning of the dimensionless couplings $f_{ij}$. In the presence of both type I and type II seesaw the neutrino mass can be written as 
\begin{equation}
\label{eq:someequation6}
        M_{\nu}=m^{II}+m^{I} 
\end{equation}             
where $m^{II} = f v_L$ is the type II seesaw contribution and $m^I = m_{LR} M_{RR}^{-1} m_{LR}^{T}$ is the type I see saw term with $m_{LR}, M_{RR}$ being Dirac and Majorana neutrino mass matrices respectively. We assume the type I seesaw to give rise to scaling neutrino mass matrix. We then introduce the type II seesaw term as a correction to the scaling neutrino mass matrix and constrain the type II seesaw term from the requirement of generating correct value of $\theta_{13}$ as well as baryon asymmetry. One interesting property of scaling is that type I seesaw can give rise to scaling neutrino mass matrix irrespective of the right handed Majorana mass matrix $M_{RR}$, if Dirac neutrino mass matrix $m_{LR}$ obeys scaling. As we discuss in section \ref{sec:numeric}, this property of scaling allows us to derive the type II seesaw correction as well as the Dirac neutrino mass matrix.

\subsection{Deviation from Scaling with Charged Lepton Correction}
The scaling neutrino mass matrix we discuss here, given by equation \eqref{eq:someequation3} predicts $m_3 = 0$ and $\theta_{13} = 0$. In the previous subsection, type II seesaw correction to scaling neutrino mass was discussed which not only can result in non-zero $\theta_{13}$ but also can give rise to non-zero $m_3$. Since, an inverted hierarchical neutrino mass pattern with $m_3=0$ is still allowed by neutrino oscillation data, one can generate non-zero $\theta_{13}$ by incorporating corrections to the leptonic mixing matrix only without affecting the scaling neutrino mass matrix. The PMNS leptonic mixing matrix is related to the diagonalizing 
matrices of neutrino and charged lepton mass matrices $U_{\nu}, U_l$ respectively, as
\begin{equation}
U_{\text{PMNS}} = U^{\dagger}_l U_{\nu}
\label{eq:someequation7}
\end{equation}
The PMNS mixing matrix can be parametrized as
\begin{equation}
U_{\text{PMNS}}=\left(\begin{array}{ccc}
c_{12}c_{13}& s_{12}c_{13}& s_{13}e^{-i\delta}\\
-s_{12}c_{23}-c_{12}s_{23}s_{13}e^{i\delta}& c_{12}c_{23}-s_{12}s_{23}s_{13}e^{i\delta} & s_{23}c_{13} \\
s_{12}s_{23}-c_{12}c_{23}s_{13}e^{i\delta} & -c_{12}s_{23}-s_{12}c_{23}s_{13}e^{i\delta}& c_{23}c_{13}
\end{array}\right) 
\label{matrixPMNS}
\end{equation}
where $c_{ij} = \cos{\theta_{ij}}, \; s_{ij} = \sin{\theta_{ij}}$ and $\delta$ is the Dirac CP phase. If $U_{\nu}$ originates from scaling neutrino mass matrix given by type I seesaw, then for diagonal charged lepton mass matrix, both the reactor mixing angle $\theta_{13}$ and the leptonic Dirac CP phase $\delta$ vanish. However, a non-trivial charged lepton mixing matrix $U_l$ can result in correct leptonic mixing matrix $U_{\text{PMNS}}$ even if $U_{\nu}$ predicts $\theta_{13}=0$. As discussed in the section on numerical analysis \ref{sec:numeric}, we constrain the charged lepton mass matrix by demanding the generation of correct $\theta_{13}$ required by neutrino oscillation data and also the correct value of $\delta_{CP}$ in order to produce correct baryon asymmetry.
\begin{figure}[h]
 \centering
\includegraphics[width=1.0\textwidth]{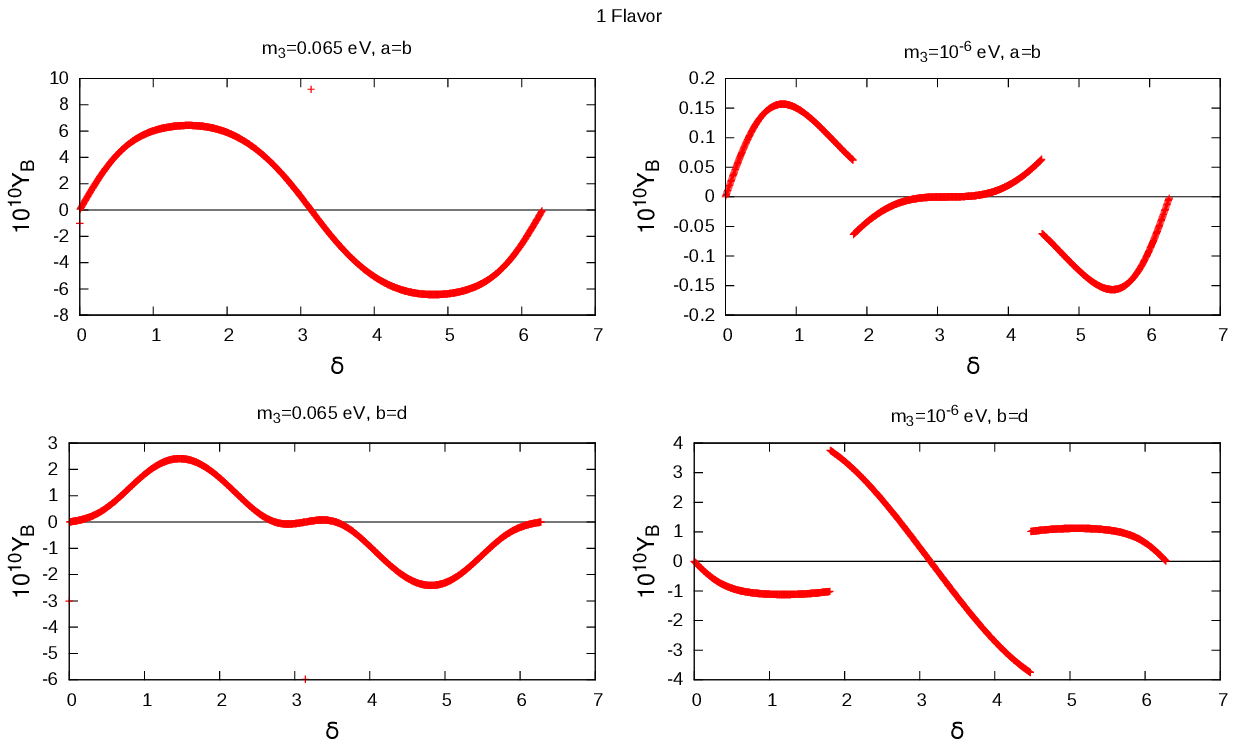}
\caption{Baryon asymmetry in one flavor regime as a function of $\delta_{CP}$ for inverted hierarchy with type II seesaw correction to scaling}
\label{fig1}
\end{figure}
\begin{figure}[h]
 \centering
\includegraphics[width=1.0\textwidth]{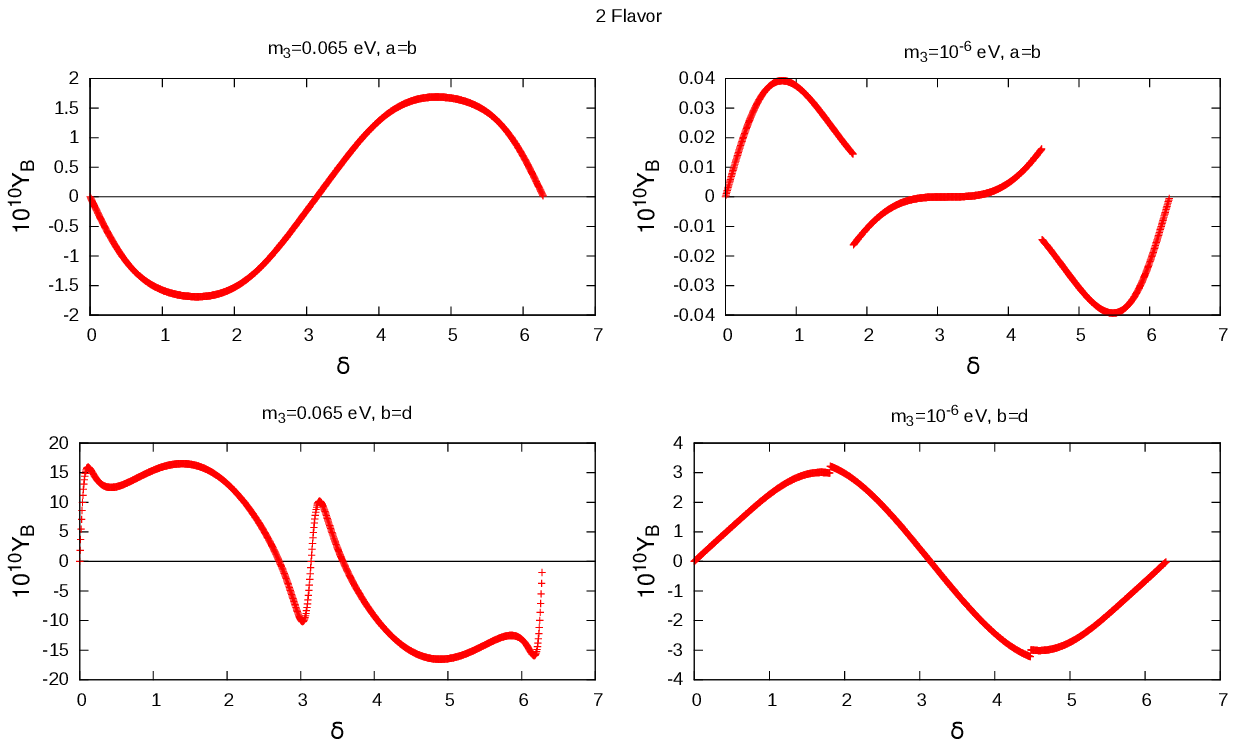}
\caption{Baryon asymmetry in two flavor regime as a function of $\delta_{CP}$ for inverted hierarchy with type II seesaw correction to scaling}
\label{fig2}
\end{figure}
\begin{figure}[h]
 \centering
\includegraphics[width=1.0\textwidth]{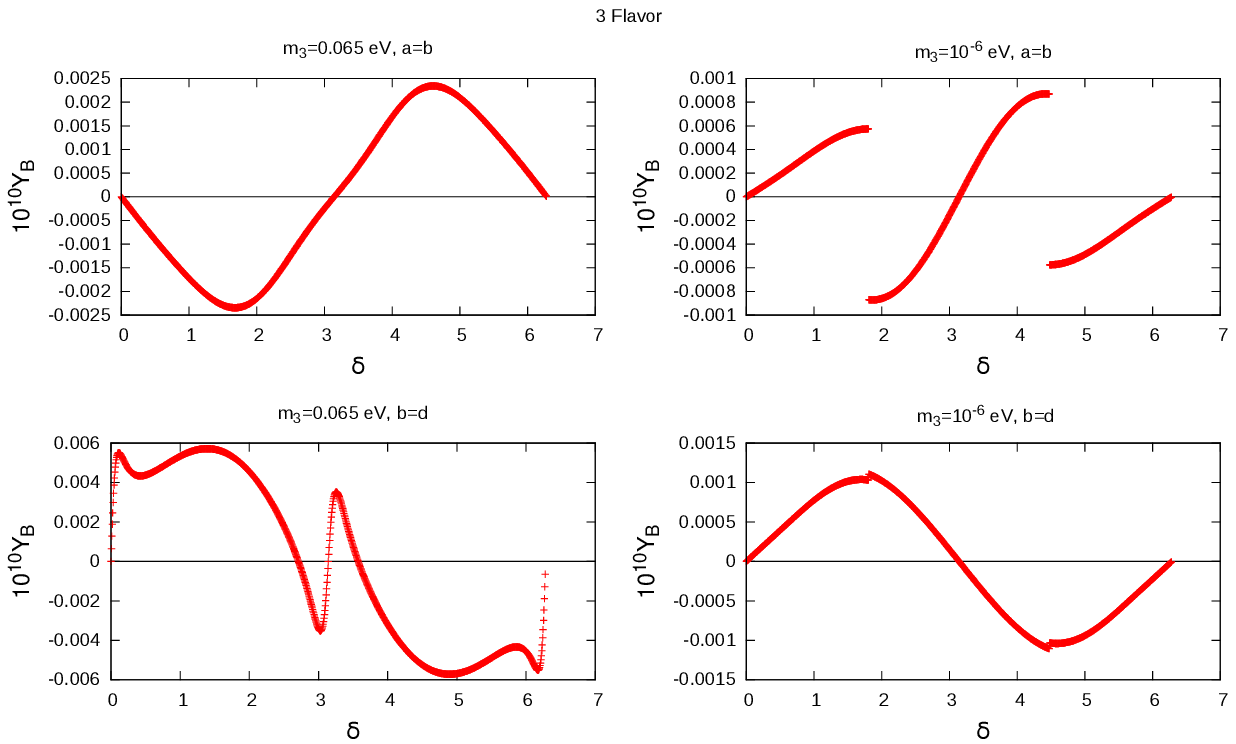}
\caption{Baryon asymmetry ratio in three flavor regime as a function of $\delta_{CP}$ for inverted hierarchy with type II seesaw correction to scaling}
\label{fig3}
\end{figure}
\begin{figure}[h]
 \centering
\includegraphics[width=1.0\textwidth]{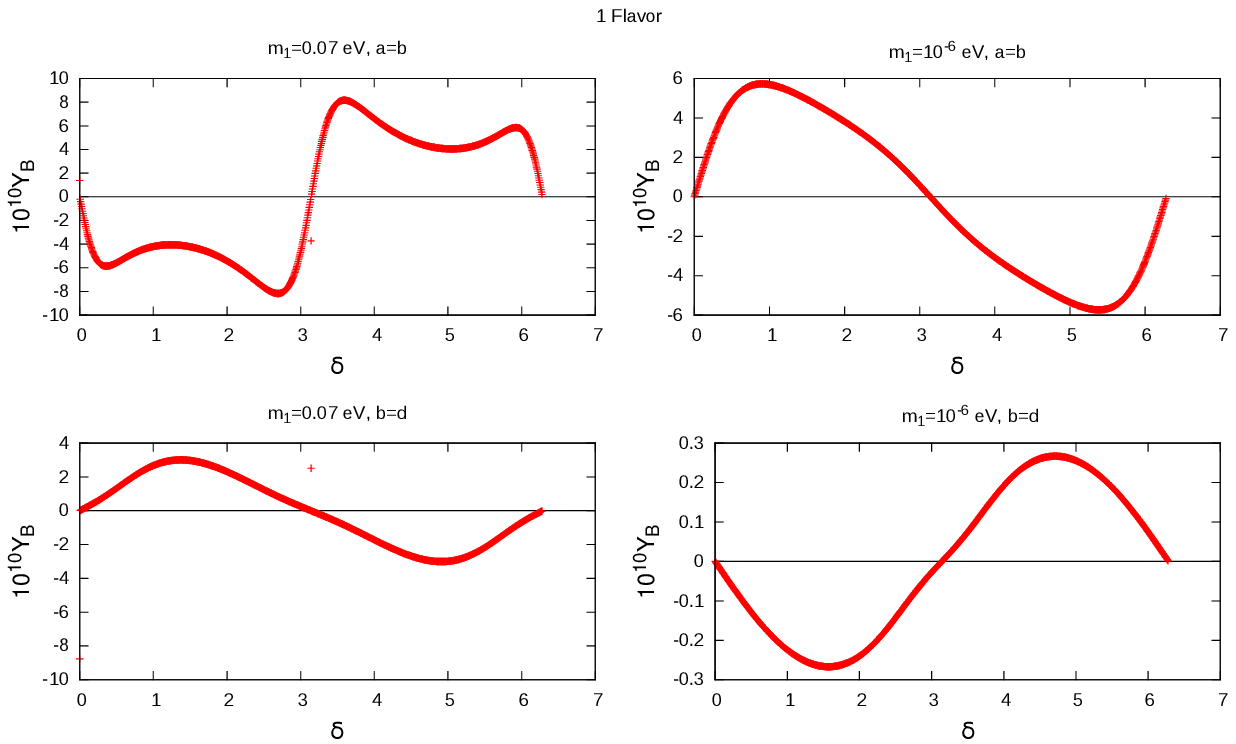}
\caption{Baryon asymmetry in one flavor regime as a function of $\delta_{CP}$ for normal hierarchy with type II seesaw correction to scaling}
\label{fig4}
\end{figure}
\begin{figure}[h]
 \centering
\includegraphics[width=1.0\textwidth]{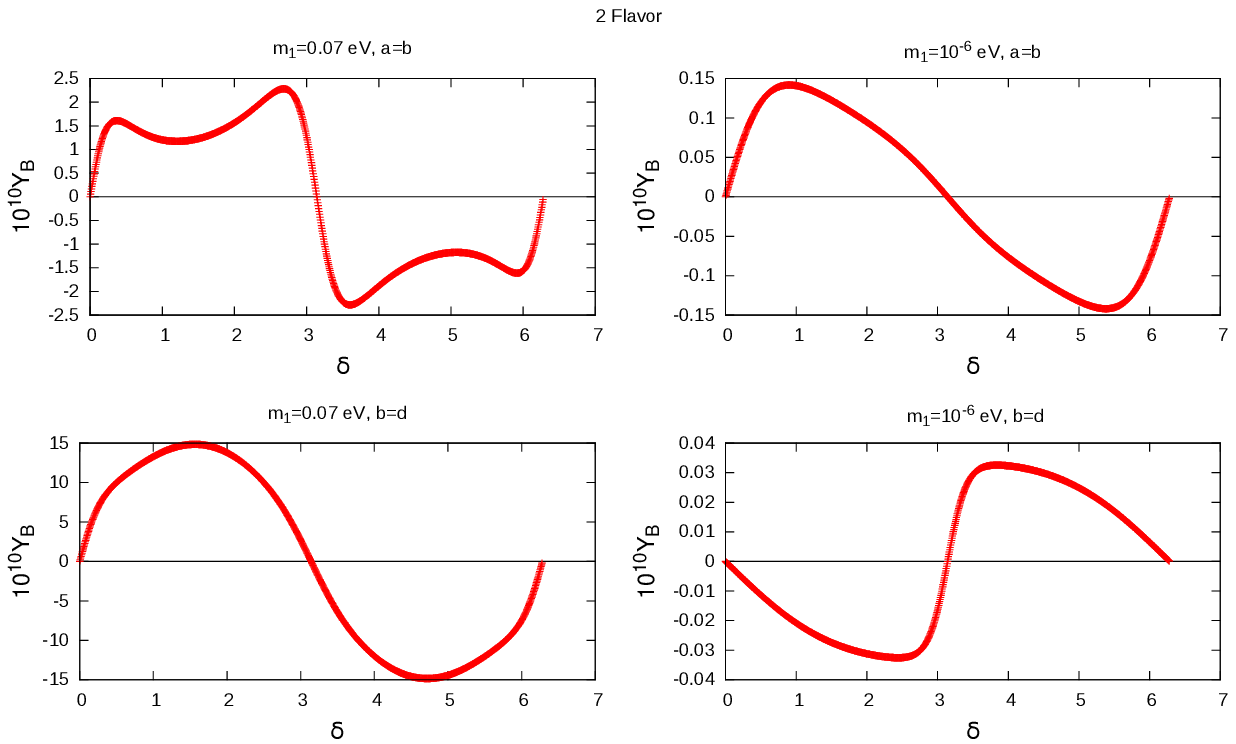}
\caption{Baryon asymmetry in two flavor regime as a function of $\delta_{CP}$ for normal hierarchy with type II seesaw correction to scaling}
\label{fig5}
\end{figure}
\begin{figure}[h]
 \centering
\includegraphics[width=1.0\textwidth]{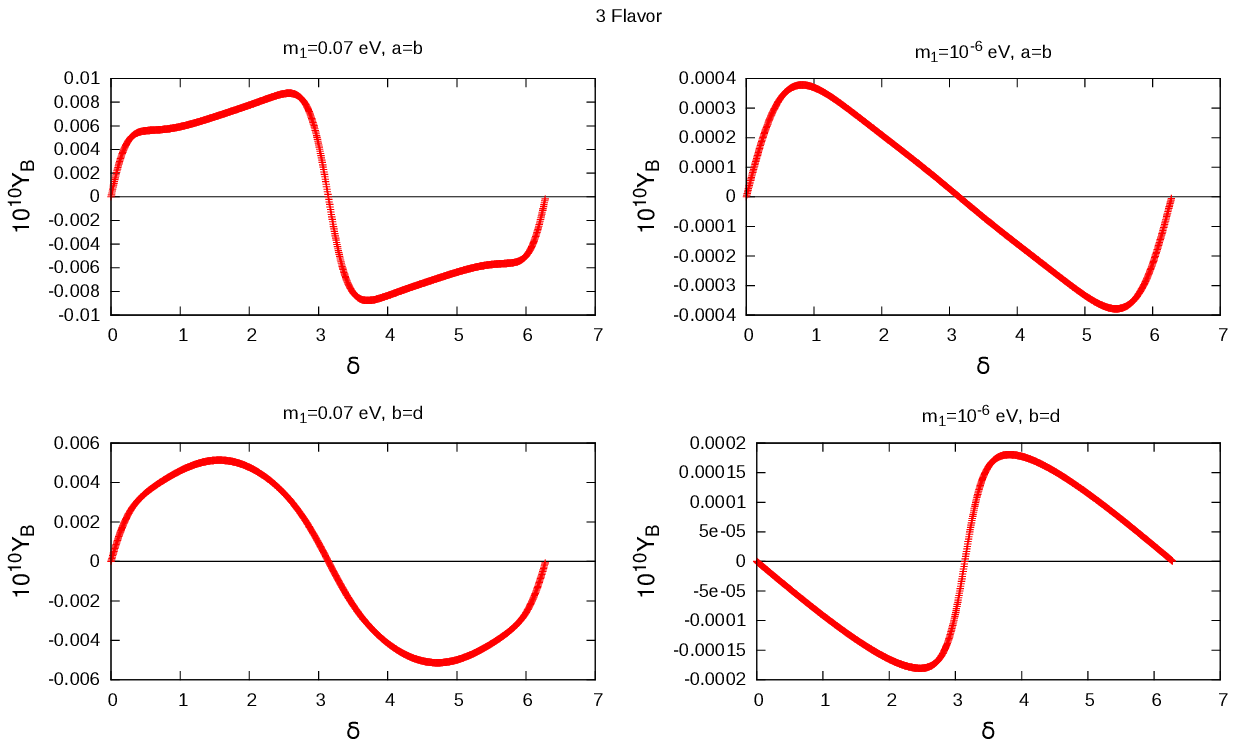}
\caption{Baryon asymmetry in three flavor regime as a function of $\delta_{CP}$ for normal hierarchy with type II seesaw correction to scaling}
\label{fig6}
\end{figure}
\begin{table}[h]
\caption{Values of $\delta_{CP}$ giving correct $Y_B$ for inverted hierarchy with Type II seesaw correction to scaling}
\begin{tabular}{|l|l|l|l|}
\hline
\multicolumn{2}{|l|}{\textbf{Model}} &\textbf{$ \delta_{CP} $ (radian) for a=b} &\textbf{$\delta_{CP}$ (radian) for b=d}  \\ \hline
\multirow{2}{*}{1 Flavor}  & $m_3=0.065$ eV  &$0.0797-0.0848,3.0316-3.0379$   & $0.6069-0.6270,2.3147-2.3329$  \\ \cline{2-4} 
                    &  $m_3=10^{-6}$ eV & --  &  $2.9009-2.9135,5.8779-5.9099$ \\ \hline
\multirow{2}{*}{2 Flavor}  & $m_3=0.065$ eV  &$3.6612-3.6963,5.9017-5.9306$  & $0.0025-0.0031,2.6778-2.6797$  \\ \cline{2-4} 
                    & $m_3=10^{-6}$ eV  & --  & $0.3537-0.3757,2.8582-2.8739$ \\ \hline
\multirow{2}{*}{3 Flavor}  & $m_3=0.065$ eV  & --  & --  \\ \cline{2-4} 
                    & $m_3=10^{-6}$ eV  & --  & --  \\ \hline
\end{tabular}
\label{table:deltaT2IH}
\end{table}
\begin{table}[h]
\caption{Values of $\delta_{CP}$ giving correct $Y_B$ for normal hierarchy with Type II seesaw correction to scaling}
\begin{tabular}{|l|l|l|l|}
\hline
\multicolumn{2}{|l|}{\textbf{Model}} &\textbf{$ \delta_{CP} $(radian) for a=b} &\textbf{$\delta_{CP}$ (radian) for b=d}  \\ \hline
\multirow{2}{*}{1 Flavor}  & $m_1=0.07$ eV  &$3.1648-3.1660,6.2561-6.2574$  & $0.3348-0.3518,2.6760-2.6998$  \\ \cline{2-4} 
                    &  $m_1=10^{-6}$ eV & $0.0659-0.0697,2.9285-2.9411$ &  -- \\ \hline
\multirow{2}{*}{2 Flavor}  & $m_1=0.07$ eV  &$0.1017-0.1086,3.0486-3.0542$   & $0.0245-0.0263,3.0963-3.0988$  \\ \cline{2-4} 
                    & $m_1=10^{-6}$ eV  & --  & --  \\ \hline
\multirow{2}{*}{3 Flavor}  & $m_1=0.07$ eV  & --  & --  \\ \cline{2-4} 
                    & $m_1=10^{-6}$ eV  & --  & --  \\ \hline
\end{tabular}
\end{table}

\section{Leptogenesis}
\label{sec:lepto}
As mentioned earlier, leptogenesis is the mechanism where a non-zero lepton asymmetry is generated by out of equilibrium, CP violating decay of a heavy particle which later gets converted into baryon asymmetry through electroweak sphaleron transitions. In a model with both type I and type II seesaw mechanisms at work, such lepton asymmetry can be generated either by the decay of the right handed neutrinos or the heavy scalar triplet. For simplicity, here we consider only the right handed neutrino decay as the source of lepton asymmetry. One can justify this assumption in those models where type I seesaw is dominating and type II seesaw is sub-leading giving rise to a Higgs triplet heavier than the lightest right handed neutrino. The lepton asymmetry from the decay of right handed neutrino into leptons 
and Higgs scalar in a model with only type I seesaw is given by
\begin{equation}
\epsilon_{N_k} = \sum_i \frac{\Gamma(N_k \rightarrow L_i +H^*)-\Gamma (N_k \rightarrow \bar{L_i}+H)}{\Gamma(N_k \rightarrow L_i +H^*)
+\Gamma (N_k \rightarrow \bar{L_i}+H)}
\end{equation}
In a hierarchical pattern of heavy right handed neutrinos, it is sufficient to consider the decay of the lightest right handed neutrino $N_1$. Following the notations of \cite{joshipura}, the lepton asymmetry arising from the decay of $N_1$ in the 
presence of type I seesaw only can be written as
\begin{eqnarray}
\epsilon^{\alpha}_1 &=& \frac{1}{8\pi v^2}\frac{1}{(m^{\dagger}_{LR}m_{LR})_{11}} \sum_{j=2,3} \text{Im}[(m^*_{LR})_{\alpha 1}
(m^{\dagger}_{LR}m_{LR})_{1j}(m_{LR})_{\alpha j}]g(x_j) \nonumber \\&& + \frac{1}{8\pi v^2}\frac{1}{(m^{\dagger}_{LR}m_{LR})_{11}} 
\sum_{j=2,3} \text{Im}[(m^*_{LR})_{\alpha 1}(m^{\dagger}_{LR}m_{LR})_{j1}(m_{LR})_{\alpha j}]\frac{1}{1-x_j}
\label{eps1}
\end{eqnarray}
where $v = 174 \; \text{GeV}$ is the vev of the Higgs doublet responsible for breaking the electroweak symmetry, $$ g(x) = \sqrt{x} 
\left ( 1+\frac{1}{1-x}-(1+x)\text{ln}\frac{1+x}{x} \right) $$and $x_j = M^2_j/M^2_1$. The second term in the expression for $\epsilon^{\alpha}_1$ 
above vanishes when summed over all the flavors $\alpha = e, \mu, \tau$. The sum over all flavors can be written as
\begin{equation}
\epsilon_1 = \frac{1}{8\pi v^2}\frac{1}{(m^{\dagger}_{LR}m_{LR})_{11}}\sum_{j=2,3} \text{Im}[(m^{\dagger}_{LR}m_{LR})^2_{1j}]g(x_j)
\label{noflavor}
\end{equation}
From the lepton asymmetry $\epsilon_1$ given by the expression above, the corresponding baryon asymmetry can be obtained by
\begin{equation}
Y_B = c \kappa \frac{\epsilon}{g_*}
\end{equation}
through sphaleron processes \cite{sphaleron} at electroweak phase transition. Here the factor $c$ is measure of the fraction of lepton asymmetry being 
converted into baryon asymmetry and is approximately equal to $-0.55$. On the other hand, $\kappa$ is the dilution factor due to wash-out process which 
erase the asymmetry generated and can be parametrized as \cite{kolbturner}
\begin{eqnarray}
-\kappa &\simeq &  \sqrt{0.1K} \text{exp}[-4/(3(0.1K)^{0.25})], \;\; \text{for} \; K  \ge 10^6 \nonumber \\
&\simeq & \frac{0.3}{K (\ln K)^{0.6}}, \;\; \text{for} \; 10 \le K \le 10^6 \nonumber \\
&\simeq & \frac{1}{2\sqrt{K^2+9}},  \;\; \text{for} \; 0 \le K \le 10.
\end{eqnarray}
where K is given as
$$ K = \frac{\Gamma_1}{H(T=M_1)} = \frac{(m^{\dagger}_{LR}m_{LR})_{11}M_1}{8\pi v^2} \frac{M_{Pl}}{1.66 \sqrt{g_*}M^2_1} $$
Here $\Gamma_1$ is the decay width of $N_1$ and $H(T=M_1)$ is the Hubble constant at temperature $T = M_1$. The factor $g_*$ is the 
effective number of relativistic degrees of freedom at temperature $T=M_1$ and is approximately $110$.

We note that the lepton asymmetry shown in equation (\ref{noflavor}) is obtained by summing over all the flavors $\alpha = e, \mu, \tau$. 
A non-vanishing lepton asymmetry is generated only when the right handed neutrino decay is out of equilibrium. Otherwise both the forward 
and the backward processes will happen at the same rate resulting in a vanishing asymmetry. Departure from equilibrium can be estimated by 
comparing the interaction rate with the expansion rate of the Universe. At very high temperatures $(T \geq 10^{12} \text{GeV})$ all charged 
lepton flavors are out of equilibrium and hence all of them behave similarly resulting in the one flavor regime. However at temperatures 
$ T < 10^{12}$ GeV $(T < 10^9 \text{GeV})$, interactions involving tau (muon) Yukawa couplings enter equilibrium and flavor effects become 
important \cite{flavorlepto}. Taking these flavor effects into account, the final baryon asymmetry is given by 
\begin{equation}
Y^{2 flavor}_B = \frac{-12}{37g^*}[\epsilon_2 \eta\left (\frac{417}{589}\tilde{m_2} \right)+\epsilon^{\tau}_1\eta\left (\frac{390}{589}
\tilde{m_{\tau}}\right )] \nonumber
\end{equation}
\begin{equation}
Y^{3 flavor}_B = \frac{-12}{37g^*}[\epsilon^e_1 \eta\left (\frac{151}{179}\tilde{m_e}\right)+ \epsilon^{\mu}_1 \eta\left (\frac{344}{537}
\tilde{m_{\mu}}\right)+\epsilon^{\tau}_1\eta\left (\frac{344}{537}\tilde{m_{\tau}} \right )] \nonumber
\end{equation}
where $\epsilon_2 = \epsilon^e_1 + \epsilon^{\mu}_1, \tilde{m_2} = \tilde{m_e}+\tilde{m_{\mu}}, \tilde{m_{\alpha}} = \frac{(m^*_{LR})_{\alpha 1} 
(m_{LR})_{\alpha 1}}{M_1}$. The function $\eta$ is given by 
$$ \eta (\tilde{m_{\alpha}}) = \left [\left ( \frac{\tilde{m_{\alpha}}}{8.25 \times 10^{-3} \text{eV}} \right )^{-1}+ \left ( \frac{0.2\times 
10^{-3} \text{eV}}{\tilde{m_{\alpha}}} \right )^{-1.16} \right ]^{-1} $$
In the presence of an additional scalar triplet, the right handed neutrino can also decay through a virtual triplet. The contribution of this diagram to lepton asymmetry can be estimated as \cite{tripletlepto}
\begin{equation} 
\epsilon^{\alpha}_{\Delta 1}=-\frac{M_1}{8\pi v^2} \frac{\sum_{j=2,3} \text{Im} [(m_{LR})_{1j}(m_{LR})_{1\alpha}(M^{II*}_{\nu})_{j\alpha}]}
{\sum_{j=2,3} \lvert (m_{LR})_{1j}\rvert^2}
\label{eps3}
\end{equation}
We use these expressions to calculate the baryon asymmetry in our numerical analysis section discussed below.

\section{Numerical analysis}
\label{sec:numeric}
We first diagonalize the scaling neutrino mass matrix \eqref{eq:someequation3} and find its eigenvalues
$$ m_1 = \frac{1}{2S^2} \left ( D+AS^2+DS^2 -\sqrt{A^2S^4-2ADS^2(1+S^2)+D^2(1+S^2)^2+4B^2S^2(1+S^2)} \right )$$ 
$$ m_2 = \frac{1}{2S^2} \left ( D+AS^2+DS^2 +\sqrt{A^2S^4-2ADS^2(1+S^2)+D^2(1+S^2)^2+4B^2S^2(1+S^2)} \right )$$ 
$$m_3 = 0$$
We numerically evaluate the four parameters $A, B, D, S$ by equating $m_1, m_2$ to two neutrino mass squared differences $\Delta m^2_{21}, \Delta m^2_{23}$ and two non-zero mixing angles to $\theta_{12}, \theta_{23}$.

Now, in the case of type II seesaw correction to scaling neutrino mixing, we assume the charged leptons mass matrix to be diagonal so that $U_{\text{PMNS}} = U_{\nu}$. Therefore, we can write \eqref{eq:someequation6} as 
        \begin{equation}
        \label{eq:someequation8}
         U_{\text {PMNS}}.m^{\text{diag}}_{\nu}.U^{T}_{\text{PMNS}}=m^{II}+m^{I}
\end{equation}        
where $m^{\text{diag}}_{\nu}$ is the diagonal neutrino mass matrix given by $m^{\text{diag}}_{\nu} 
= \text{diag}(m_1, \sqrt{m^2_1+\Delta m_{21}^2}, \sqrt{m_1^2+\Delta m_{31}^2})$ for normal hierarchy and $m^{\text{diag}}_{\nu} = \text{diag}(\sqrt{m_3^2+\Delta m_{23}^2-\Delta m_{21}^2}, 
\sqrt{m_3^2+\Delta m_{23}^2}, m_3)$ for inverted hierarchy. The type I seesaw mass matrix $ m^{I}$ always gives inverted hierarchy whereas $m^{\text{diag}}_{\nu}$ can give either normal or inverted hierarchy depending on the type II seesaw contribution $m^{II}$. In the minimal extension of the standard model with type I and type II seesaw mechanisms, the type I seesaw term depends upon $m_{LR}$ and $M_{RR}$ whereas type II seesaw depends upon the vev of the neutral component of Higgs triplet. Since $m_{LR}, M_{RR}$ as well as the type II seesaw term can be chosen by hand, such a framework is difficult to constrain due to too many number of free parameters. However, in a specific class of models called left right symmetric models (LRSM) \cite{lrsm}, the type II seesaw term is directly proportional to $M_{RR}$ thereby decreasing the number of free parameters compared to the minimal extension. Another reason for choosing the framework of LRSM is that here we can find the right handed Majorana mass matrix $M_{RR}$ from the type II seesaw perturbation. However, for a given Dirac neutrino mass matrix $m_{LR}$, one can not find $M_{RR}$ from the type I seesaw formula alone as the inverse of type I seesaw mass matrix does not exist due to its scaling property $(m_3=0)$. In LRSM we can write equation \eqref{eq:someequation8} as
 \begin{equation}
 \label{eq:someequation9}
        U_{\text {PMNS}}.m^{\text{diag}}_{\nu}.U^{T}_{\text{PMNS}}={\gamma}\left(\frac{M_W}{v_R}\right)^2M_{RR}+m^{I}
\end{equation}      
where $\gamma$ is a dimensionless parameter, $M_W$ is the W boson mass and $v_R$ is the scale of left right symmetry breaking. Since $m^I$ has been numerically evaluated as the leading order scaling neutrino mass matrix, type II contribution can now be evaluated as a function of leptonic Dirac CP phase $\delta_{CP}$ and the lightest neutrino mass $m_1$ (NH), $m_3$ (IH), the two unknowns on the left hand side of the above equation. It should be noted that, we have omitted the Majorana phases in this discussion. After determining the type II seesaw term and hence $M_{RR}$, we use it in the type I seesaw term to find out the Dirac neutrino mass matrix $m_{LR}$. Here we use the already mentioned special property of scaling neutrino mass matrix originating from type I seesaw: if Dirac neutrino mass matrix $m_{LR}$ obeys scaling, then $m^I$ obeys scaling irrespective of the structure of $M_{RR}$. Therefore, we use the scaling Dirac neutrino mass matrix given by 
\[
m_{LR}=
  \begin{pmatrix}
    a & b & \frac{b}{c} \\        b & d &\frac{d}{c} \\ \frac{b}{c} &\frac{d}{c} &\frac{d}{c^{2}}      \end{pmatrix}
\]
We use the above $m_{LR}$ and the already derived right handed Majorana mass matrix $M_{RR}$ in the type I seesaw formula and equate it to the scaling neutrino mass matrix evaluted numerically earlier. 

We use two different choices of lightest neutrino mass in order to show the effect of hierarchy. For inverted hierarchy we take $m_3 = 0.065$ eV, $10^{-6}$ eV and for normal hierarchy we take $m_1 = 0.07$ eV, $10^{-6}$ eV. After choosing the lightest neutrino mass, the only undetermined parameters in the equation \eqref{eq:someequation9} are $\delta_{CP}, \gamma$ and $v_R$. Choosing generic order one coupling $\gamma$, one can now write $M_{RR}$ in terms of $\delta_{CP}$ and $v_R$. We choose the left right symmetry breaking scale $v_R$ in a way which keeps the lightest right handed neutrino in the appropriate flavor regime of leptogenesis. After we find $M_{RR}$ in terms of $\delta_{CP}$, we use it in the type I seesaw formula with the $m_{LR}$ obeying scaling as shown above. To simplify the numerical calculation further, we assume equality between some parameters in $m_{LR}$: $a=b$ and $b=d$. The other choice $a=d$ does not give us any solution. We also do not assume equality of the parameter $c$ with others as $c$ can be found independently of $a, b, d$ when we equate the type I seesaw formula with the numerically evaluated type I seesaw mass matrix of scaling type. The numerical form of the right handed neutrino mass matrix $M_{RR}$ for all the cases discussed are shown in table \ref{table:MRRT21}, \ref{table:MRRT22}, \ref{table:MRRT23} and \ref{table:MRRT24}. Similarly, the Dirac neutrino mass matrices are listed in table \ref{table:mLRT21}, \ref{table:mLRT22}, \ref{table:mLRT23}, \ref{table:mLRT24}, \ref{table:mLRT25}, \ref{table:mLRT26}, \ref{table:mLRT27} and \ref{table:mLRT28}. Although they are, in general, complicated functions of $\delta_{CP}$, we have used a specific value of $\delta_{CP} = \pi/2$ to show their compact numerical form. To calculate the baryon asymmetry, however, we vary $\delta_{CP}$ continuously and show the variation of Baryon asymmetry in figure \ref{fig1}, \ref{fig2}, \ref{fig3}, \ref{fig4}, \ref{fig5} and \ref{fig6}. It should be noted that the type II seesaw corrections to scaling have been considered within the framework of LRSM where $SU(2)_R$ gauge interactions can give rise to sizeable wash-out effects erasing the asymmetry produced. As noted in \cite{washoutLR}, such wash-out effects can be neglected by choosing a high value of $v_R$ such that $M_1/v_R < 10^{-2}$ is satisfied. \\
\begin{table}[h]
\caption{Charge lepton diagonalizing matrix for $\delta_{CP}=\frac{\pi}{2}$}
\begin{center}

\resizebox{\textwidth}{!}{%
\begin{tabular}{ | c| c|  }
\hline
 \text{ $U_l$  }\\
\hline

\addlinespace[1.5ex]
\hline
 
$ \begin{pmatrix}0.389343 - 0.108352 i&0.0141579 + 0.0381397 i &-0.933412 + 0.0458869 i\\-0.412832 + 0.0492513 i &0.881799 - 0.0404407 i&-0.313989 - 0.0486552 i\\0.925292 + 0.059714 i &0.3356 + 0.0906409 i &0.228991 + 0.109052 i \end{pmatrix}$\\
\hline

\end{tabular}
}
\end{center}
\label{table:Ul}
\end{table}
\begin{table}[h]
\caption{$M_{RR}$ (in GeV) with charged lepton correction to scaling}
\begin{center}

\begin{tabular}{ |c| c| c|  }
\hline\text{Model} & \text{$M_{RR}$ }(GeV)\\
\hline

\addlinespace[0.5ex]
\hline
 1 Flavor &
$\left(\begin{array}{ccc}
1\times 10^{13} & 0 & 0 \\
\ 0 & 2 \times10^{13}& 0\\
\ 0 &0 & 3\times10^{13}\\
\end{array}\right)$ \\
\hline
\addlinespace[0.5ex]
\hline
 2 Flavor &
$\left(\begin{array}{ccc}
\ 1\times10^{10} & 0 & 0 \\
\ 0 &1\times 10^{13}& 0\\
\ 0 & 0& 3\times 10^{16}\\
\end{array}\right)$\\
\hline
\addlinespace[0.5ex]
\hline
 3 Flavor &
$\left(\begin{array}{ccc}
\ 1\times10^8 & 0 & 0 \\
\ 0 & 1\times10^{9}& 0\\
\ 0 & 0& 1\times10^{15}\\
\end{array}\right)$\\
\hline
\end{tabular}

\end{center}
\label{table:MRRCL}
\end{table}
\begin{table}[h]
\caption{$m_{LR}$ (in GeV) for $\delta_{CP}=\frac{\pi}{2}$ with charged lepton correction to scaling}
\begin{center}

\resizebox{\textwidth}{!}{%
\begin{tabular}{ |c| c| c|  }
\hline
 & \textbf{NH $m_1=10^{-6}$ eV}\\
\hline

\addlinespace[1.5ex]
\hline
 \text{$1$ Flavor} &
$ \begin{pmatrix}-18.6984 + 6.60236  i&12.7243 - 1.50446 i&-8.96417 - 11.8754 i\\11.7542 - 1.7535 i&-8.01219 + 0.437544 i&5.72898 + 3.02685 i\\-4.25822 - 10.3367 i&2.93615 + 2.10775 i&-2.36918 + 19.5204  i \end{pmatrix}$\\
\hline
\addlinespace[2.5ex]
\hline
  \text{$2$ Flavor} &
$ \begin{pmatrix}-4.11131 + 36.0857 i&4.01777 - 17.0146 i&-8.06708 - 20.2467 i\\0.42871 - 17.3158 i&-1.20776 + 8.25435 i &4.77669 + 9.26783  i\\9.83461 - 18.7204 i&-5.67658 + 8.37772 i&-0.331627 + 12.7398 i\end{pmatrix}$\\
\hline
\addlinespace[2.5ex]
\hline
 \text{$3$ Flavor} &
$ \begin{pmatrix}0.175984 - 0.191103  i&-0.111308 + 0.0777845 i&0.0423717 + 0.169387 i\\-0.0896843 + 0.0881861 i&0.057542 - 0.0366476 i&-0.0257287 - 0.074506 i\\-0.0647843 + 0.116022 i&0.0369503 - 0.0433691 i& 0.00475925 - 0.121585 i\end{pmatrix}$\\
\hline
\end{tabular}
}
\end{center}
\label{table:mLRCL}
\end{table}
\begin{table}[h]
\caption{Values of $\delta_{CP}$ giving correct $Y_B$ with charge lepton correction to scaling}
\begin{tabular}{|l|l|}
\hline
\textbf{Model} & \textbf{$\delta_{CP}$ (radian)} \\ \hline
 1 Flavor & $2.2003-2.2072,5.2552-5.2734$ \\ \hline
 2 Flavor & $3.2440-3.2452,6.2586-6.2624$ \\ \hline
 3 Flavor & $4.1010-4.1223,6.0136-6.0670$ \\ \hline
\end{tabular}
\label{table:deltaCL}
\end{table}

\begin{figure}[h]
 \centering
\includegraphics[width=1.0\textwidth]{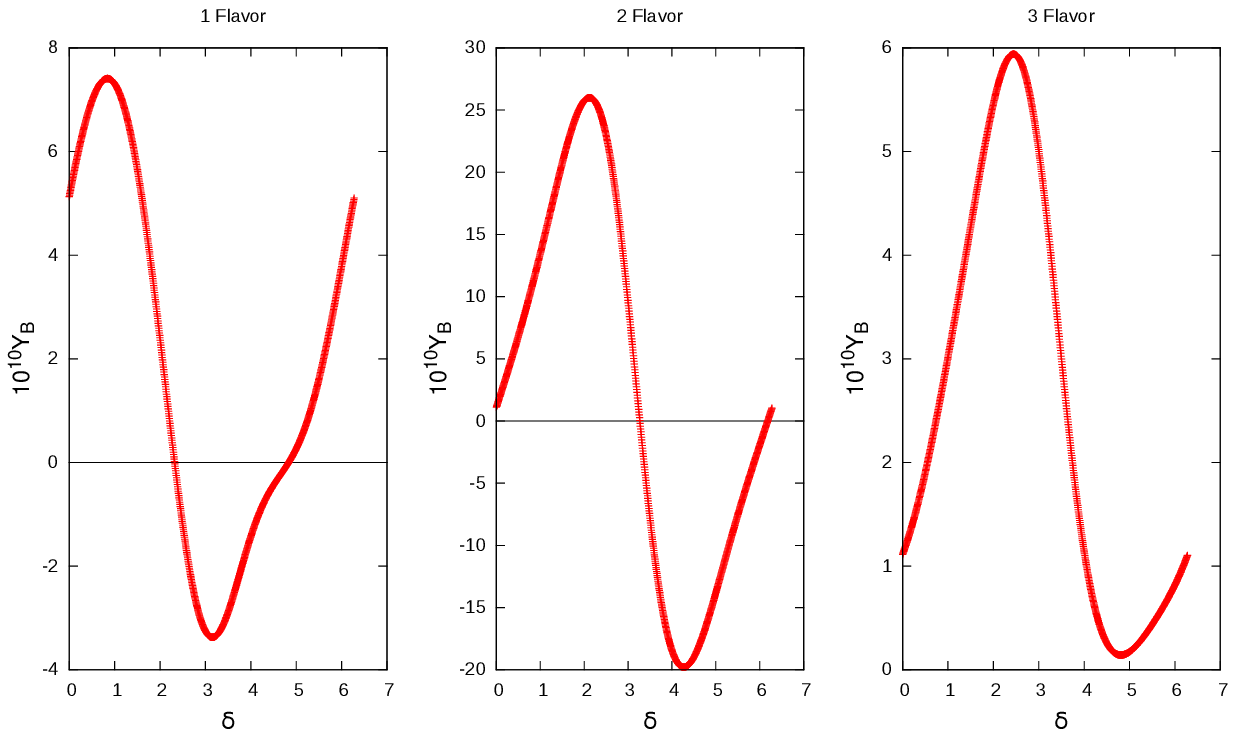}
\caption{Baryon asymmetry in one, two and three flavor regimes as a function of $\delta_{CP}$  with charged lepton correction to scaling}
\label{fig7}
\end{figure}
In the second mechanism we adopt to give correction to the scaling neutrino mixing, we do not add corrections to the neutrino mass matrix originating from type I seesaw, but incorporate corrections to the neutrino mixing matrix originating from charged lepton mixing. Diagonalizing the scaling neutrino mass matrix from type I seesaw matrix gives $U_{\nu}$ which is related to the leptonic mixing matrix $U_{\text{PMNS}}$ through the charged lepton diagonalizing matrix $U_l$. We first numerically evaluate $U_{\nu}$ by using the best fit values of neutrino mass squared differences and two mixing angles $\theta_{12}, \theta_{23}$. We also substitute the best fit values of neutrino mixing angles in PMNS mixing matrix \eqref{matrixPMNS} and then compute the charged lepton diagonalizing matrix as 
$$ U_l = U_{\nu} U^{\dagger}_{\text{PMNS}}$$
We keep the Dirac CP phase $\delta_{CP}$ as free parameter so that $U_{l}$ is a function of it. The numerical form of $U_l$ for $\delta_{CP}=\pi/2$ is shown in table \ref{table:Ul}. Assuming the same diagonalizing matrix of charged leptons and Dirac neutrino mass matrix (a generic case in grand unified theories operating at high scale), we can write down the modified Dirac neutrino mass matrix as 
       \begin{equation}
       \label{eq:someequation11}
        m_{LR}=U_{l}.m^{0}_{LR}.U^{\dagger}_{l}
\end{equation}       
where $m^0_{LR}$ is the scaling type Dirac neutrino mass matrix we choose earlier. We choose a diagonal form of $M_{RR}$ while keeping the lightest right handed neutrino mass $M_1$ in the appropriate flavor regime of leptogenesis and varying the heavier right handed neutrino masses $M_2, M_3$ between $M_1$ and the grand unified theory scale $M_{\text{GUT}} \sim 10^{16}$ GeV. For such a choice of $M_{RR}$, we numerically evaluate the parameters in $m^0_{LR}$ by equating the type I seesaw term $\left (m^0_{LR}\right )M^{-1}_{RR}\left (m^0_{LR} \right )^T$ to the numerically fitted type I scaling neutrino mass matrix. The numerical values of $M_{RR}$ which give baryon asymmetry closest to the observed in each flavor regime are shown in table \ref{table:MRRCL}. The numerical form of Dirac neutrino mass matrices in each flavor regime for $\delta_{CP}=\pi/2$ are shown in table \ref{table:mLRCL}. The predictions for baryon asymmetry as a function of $\delta_{CP}$ are shown in figure \ref{fig7}. The values of Dirac CP phase which give rise to correct baryon asymmetry are listed in table \ref{table:deltaCL}.

\section{Results and Conclusion}
\label{sec:conclude}
We have studied a specific type of neutrino mass matrix based on the idea of strong scaling ansatz where the ratio of neutrino mass matrix elements belonging to two different columns are equal. Out of three such possibilities, we focus on a particular scaling neutrino mass matrix which predicts zero values of reactor mixing angle $\theta_{13}$. This choice was motivated by several recent works where the leading order neutrino mass matrix obeying certain symmetries predict $\theta_{13}=0$ and suitable corrections to the neutrino mass matrix or leptonic mixing matrix give rise to small but non-zero values of $\theta_{13}$. In this work, we have assumed type I seesaw to give rise to scaling neutrino mass matrix which (in the diagonal charged lepton basis) gives $\theta_{13}=0$ and inverted hiearchical mass pattern with $m_3=0$. Then we consider two different possible corrections to scaling: one with type II seesaw which gives rise to deviations from both $\theta_{13}=0$ and $m_3=0$, the other with charged lepton correction which gives non-zero $\theta_{13}$ while keeping $m_3=0$. We also assume both the corrections to give rise to non-trivial Dirac CP phase $\delta_{CP}$ as well. In both the cases, we first numerically evaluate the type I seesaw scaling neutrino mass matrix by using the best fit values of neutrino mass squared differences and two mixing angles: solar and atmospheric. We then calculate the necessary corrections to scaling neutrino mass and mixing by keeping the Dirac CP phase as free parameter. We further constrain the Dirac CP phase by calculating the baryon asymmetry through the mechanism of leptogenesis and comparing with the observed Baryon asymmetry. The important results we have obtained in the case of type II seesaw correction to scaling can be summarized as:
\begin{itemize}
\item Type II seesaw correction to scaling neutrino mass matrix with $\theta_{13}=0, m_3=0$ can result in both normal as well as inverted hierarchy with non-zero $\theta_{13}$ as well as non-trivial Dirac CP phase $\delta_{CP}$.
\item For inverted hierarchy with $a=b$ that is $m_{LR}(11) = m_{LR}(12)$, correct values of baryon asymmetry is obtained through the mechanism of leptogenesis only when the lightest neutrino mass $m_3$ is of same order as the heavier ones $m_1, m_2$. 
\item For inverted hierarchy with $b=d$ that is $m_{LR}(12) = m_{LR}(22)$, both large and mild hierarchy among neutrino masses give rise to correct baryon asymmetry through leptogenesis.
\item For normal hierarchy with $m_{LR}(11) = m_{LR}(12)$ , both large and mild hierarchy among neutrino masses can give rise to correct baryon asymmetry in the one flavor regime. In the two flavor regime however, the lightest neutrino mass $m_3$ should be of same order as $m_2, m_3$ to give correct baryon asymmetry.
\item For normal hierarchy with $m_{LR}(12) = m_{LR}(22)$, the lightest neutrino mass $m_3$ should be of same order as $m_2, m_3$ to give correct baryon asymmetry in both one and two flavor regimes.
\item Observed baryon asymmetry can not be generated in the three flavor regime of leptogenesis in this framework.
\end{itemize}
Similarly, the important results in the case of charged lepton correction to scaling are:
\begin{itemize}
\item Charged lepton correction to scaling neutrino mixing predicts only inverted hierarchy with $m_3=0$, but gives rise to correct values of $\theta_{13}$ and non-trivial $\delta_{CP}$.
\item Correct baryon asymmetry can be obtained through leptogenesis for one, two and three flavor regimes if $\delta_{CP}$ is restricted to certain range of values.
\end{itemize}
Since the Dirac CP phase is restricted in all these cases discussed, from the demand of producing the correct baryon asymmetry, future determination of $\delta_{CP}$ should be able to shed some light on these scenarios. Future experiments may however, measure a different value of $\delta_{CP}$ than the ones which give correct baryon asymmetry through the mechanism of leptogenesis in the models we have studied here. This will by no means rule out the neutrino mass models based on strong scaling ansatz we discuss, but will only hint at a different source of baryon asymmetry than the one discussed in our work. Similarly, determination of neutrino mass hierarchy in neutrino oscillation experiments will further constrain the models and only charged lepton correction to scaling may not be sufficient to reproduce the correct neutrino data if inverted hierarchy gets disfavored by experiments. From theoretical point of view, such scaling neutrino mass matrix can find a dynamical origin within discrete flavor symmetry models as pointed out by \cite{PLB644}. Since scaling is not affected by renormalization group running, additional physics are required in order to produce correct low energy neutrino oscillation data. Undisturbed by such running effects, scaling can be valid all the way from grand unified theory scale down to the TeV scale, where new physics affects like Higgs triplet in type II seesaw can give rise to the necessary correction to scaling neutrino mass matrix. Although we have studied only one particular type of scaling neutrino mass matrix giving $\theta_{13}=0, m_3=0$, the other two possibile scaling mass matrices could also give rise to correct neutrino phenomenology if suitable corrections are incorporated, which is left for our future studies.

\section{Acknowledgement}
The work of M. K. Das is partially supported by the grant no. 42-790/2013(SR) from University Grants Commission, Government of India.

\end{document}